\documentclass[longbibliography, aps, pra, amsmath, amssymb, amsfonts, twocolumn,
superscriptaddress, footinbib]{revtex4-2}
\pdfoutput=1

\usepackage{graphicx}
\usepackage{subfigure}
\usepackage{dcolumn}
\usepackage{bm}
\usepackage{bbm}
\usepackage{amsthm}
\usepackage{mathtools}
\usepackage{physics}
\usepackage{binarytree}
\usepackage{tikz}
\usepackage{verbatim}
\usepackage{pgfplots}
\usepackage{placeins}
\usetikzlibrary{decorations.pathreplacing}
\usetikzlibrary{arrows.meta}
\usetikzlibrary{graphs}

\usepackage[colorlinks, linkcolor=red, anchorcolor=blue, citecolor=green]{hyperref}

\pgfplotsset{compat=1.16}
\begin{document}

\title[Topological correlation: anyonic states cannot be determined
by local operations and classical communication]
{Topological correlation: anyonic states cannot be determined
by local operations and classical communication}

\author{Cheng-Qian Xu}

\affiliation{Institute of Physics, Beijing National Laboratory for Condensed
  Matter Physics,\\Chinese Academy of Sciences, Beijing 100190, China}

\affiliation{School of Physical Sciences, University of Chinese Academy of
  Sciences, Beijing 100049, China}

\author{D. L. Zhou} \email[]{zhoudl72@iphy.ac.cn}

\affiliation{Institute of Physics, Beijing National Laboratory for Condensed
  Matter Physics,\\Chinese Academy of Sciences, Beijing 100190, China}

\affiliation{School of Physical Sciences, University of Chinese Academy of
  Sciences, Beijing 100049, China}

\date{\today}

\begin{abstract}
    Anyonic system not only has potential applications in the construction of
    topological quantum computer, but also presents a unique property known as
    topological entanglement entropy in quantum many-body systems.
    How to understand topological entanglement entropy is one of the most concerned
    problems for physicists.
    For an anyonic bipartite system, we define an operational measure of topological
    correlation based on the principle of
    maximal entropy, where the topological correlation is the information that cannot be
    accessed by local operations constrained by anyonic superselection rules and
    classical communication. This measure can be extended to measure non-local resources
    of other compound quantum systems in the presence of superselection rules.
    For a given anyonic bipartite state with maximal rank, we prove that its topological
    correlation is equal to its entropy of anyonic charge entanglement that
    has been shown in the literature to be able to derive topological entanglement
    entropy.
    This measure provides a more refined classification of correlations in a multipartite
    system with superselection rules and an illuminating approach to topological phase
    classification.
\end{abstract}

                              
\maketitle

\emph{Introduction}--
Anyon~\cite{PhysRevLett.49.957, PhysRevLett.48.1144}, as a quasi-particle different
from fermions and bosons in two-dimensional systems, attracts the attention of
theorists and experimentalists by virtue of its potential in building a topological
quantum computer~\cite{kitaev2003fault}. Since its peculiar structure of Hilbert
space due to superselection rules (SSR)~\cite{PhysRevA.69.052326}
and fusion algebra~\cite{KITAEV20062, categoryphysics}, systems with anyons exhibit
distinctive properties. When a two-dimensional system stays in a topologically
ordered phase, the entanglement entropy of its ground state contains a constant
associated with anyons, which is called topological entanglement
entropy~\cite{kitaev2006topological, levin2006detecting} (TEE). This peculiar
property opens the door to characterizing a class of topological phases with
long-range entanglement~\cite{PhysRevB.82.155138}.

A lot of analytical and numerical work has attempted to elucidate TEE so far, However,
we do not yet have a definitive understanding of it. On the one hand, it's believed that
the TEE is associated with irreducible multiparty
correlations~\citep{PhysRevLett.89.207901, PhysRevLett.101.180505, Chen_2015, Liu_2016}.
The latter is a measure based on the principle of maximal
entropy~\cite{PhysRev.106.620, PhysRev.108.171},
which describes the genuine correlations that belong to the total system but cannot be
obtained from the information of the local subsystems. And it has be
proved that, under some assumptions, these two measures TEE and irreducible
multiparty correlations coincide~\citep{PhysRevA.93.022317}.
On the other hand, It has been shown that the TEE can be derived from the formal
definition of the entropy of anyonic charge
entanglement~\cite{bonderson2017anyonic} (ACE) based on anyon model~\cite{KITAEV20062}.
The ACE characterizes the correlations arising from anyonic charge lines connecting
two subsystems in a bipartite system.
The above two measures give an explanation of TEE from two different perspectives.
In this letter, we establish a connection between two ideas above.

TEE is the property of ground states of many-body systems in which the abundance of
symmetries gives a variety of characteristics, such as SSR that forbids the superposition
of quantum states that do not belong to the same class~\cite{PhysRev.88.101}.
From an operational perspective, it is realized that SSR enforces additional
restrictions on local operations and gives rise to non-local resources in a bipartite
setting~\cite{PhysRevLett.91.010404, PhysRevLett.92.087904, PhysRevLett.91.097903}.
It is reasonable to speculate that this non-local resource is associated with TEE. However,
Most of the related work has studied SSR arising from compact symmetry
groups~\cite{PhysRevLett.92.087904, PhysRevLett.91.097903, deGroot_2020, PhysRevLett.130.150201},
which can not be directly used to study SSR in anyonic systems.

In this letter, we extend the method
in Ref.~\cite{PhysRevLett.89.207901, PhysRevLett.101.180505} to define a measure
of the non-local degrees of freedom in anyonic systems which we call topological
correlations. This measure can be regarded as an extension from systems with SSR arising
from compact symmetry groups to systems with more general SSR.
Specifically, for a bipartite system in state $\rho^{AB}$, we allow two players,
Alice and Bob, to obtain information about the system through their respective
local operations and classical communication (LOCC) constrained by SSR, and let them
use the obtained information to infer the quantum state $\sigma_m(\rho^{AB})$ of the
total system based on the principle of maximum entropy. The topological correlation
is defined as the difference between the entropy of the two quantum states $\rho^{AB}$
and $\sigma_m(\rho^{AB})$. We find that conventional bipartite quantum states with no
superselection rules do not have the topological correlation. In other words, they can be
determined by Alice and Bob through LOCC while anyonic bipartite states cannot.
Furthermore we give the analytical form of the inferred
state $\tilde{\sigma}_m(\tilde{\rho}^{AB})$ for anyonic bipartite
state $\tilde{\rho}^{AB}$ with maximal rank, and prove that its topological
correlations is equivalent to its ACE.

\emph{Anyon model}--
Here we give a cursory review of anyon models~\cite{KITAEV20062}. An anyon model $\mathcal{C}$
consists of a finite collection of
topological charges ${\rm ob}(\mathcal{C}) = \left\{ 1, a, b, \cdots \right\}$
obeying fusion rules:
 $a \times b = \sum_c N_{ab}^c c$,
where nonnegative integer $N_{ab}^c$ named fusion coefficient represents the number
of ways for charges $a$ and $b$ fuse into charge $c$.
Charge $1$ in set ${\rm ob}(\mathcal{C})$ denotes the vacuum. And every charge $a$
fuses with the vacuum resulting in itself. If fusion result of charge $a$ with
any charge $b$ is unique, then anyon carried charge $a$ is called abelian anyon
otherwise non-abelian anyon.

Based on the fusion rules above, we can build the Hilbert space for an anyonic system.
Here we consider two anyons $a$ and $b$ with total charge $c$. The corresponding
anyonic Hilbert space is the space $V^{ab}_c$, which is spanned by
vectors $\ket{a, b; c, \mu}$, where $\mu = 1, 2, \cdots, N^{ab}_c$. We can
also define dual space $V_{ab}^c$ spanned by dual vectors $\bra{a, b; c, \mu}$,
In the diagrammatic representation, these states above can be rewritten as
\begin{align}
    & \ket{a, b; c, \mu} = \left( \frac{d_c}{d_a d_b} \right)^{1/4}
        \begin{tikzpicture}[baseline]
            \draw (-0.5,0.5) -- (0,0) node[pos = 0, above]{$a$} -- (0.5,0.5)
                node[pos = 1, above]{$b$};
            \draw (0,0) -- (0,-0.5) node[pos = 1, below]{$c$};
            \draw (0,0) node[right]{$\mu$};
        \end{tikzpicture}, \nonumber \\
    & \bra{a, b; c, \mu} = \left( \frac{d_c}{d_a d_b} \right)^{1/4}
    \begin{tikzpicture}[baseline]
        \draw (-0.5,-0.5) -- (0,0) node[pos = 0, below]{$a$} -- (0.5,-0.5)
            node[pos = 1, below]{$b$};
        \draw (0,0) -- (0,0.5) node[pos = 1, above]{$c$};
        \draw (0,0) node[right]{$\mu$};
    \end{tikzpicture}, \nonumber
\end{align}
where $d_a$ is the quantum dimension of charge $a$, and each anyon is associated
with an oriented (the arrow is omitted here) line going up from bottom.
For more than two anyons, we need to specify the order of fusion, since
different orders will give different bases that describe the same anyonic Hilbert
space. These different bases can be transformed in to each other by natural
isomorphic transformations named $F$ matrices.

In addition to fusion rules, anyon model also meets braiding rules. Specifically,
exchanging two anyons will give a natural isomorphic transformation,
called $R$ matrix, acting on the system.

\emph{Information-theoretic approach to topological correlation}--
This part we consider a quantum information task called local bipartite
quantum state tomography (LBQST), whose purpose is to get Alice and Bob to
determine bipartite quantum state $\rho^{AB}$ of quantum system with SSR as best as
possible through joint measurements.
Specifically, for given many copies of bipartite quantum states $\rho^{AB}$,
Alice and Bob have states $\rho^A$ and $\rho^B$, respectively,
where $\rho^A = {\rm Tr}_B\left[ \rho^{AB} \right]$
and $\rho^B = {\rm Tr}_A \left[ \rho^{AB} \right]$. Alice and Bob perform quantum
state tomography by joint measurements, i.e., they measure the quantum states
they have through local operations constrained by SSR and classical communication,
to determine state $\rho^{AB}$.

For conventional bipartite quantum states without SSR, Alice and Bob can accurately
determine them in principle. Without loss of generality, let's take a two-qubit state as
an example. Since any two-qubit state $\rho^{AB}$ can be formulated as
 $\rho^{AB} = \frac{1}{4} \sum_{\alpha, \beta} c_{\alpha\beta} \sigma_\alpha
 \otimes \sigma_\beta$,
where $\sigma_0 = I$ is the identity matrix, $\sigma_\alpha$, $\alpha =1,2,3$, are the Pauli
matrices forming the set of generators of SU($2$) group, $c_{\alpha\beta}$ is factor,
and $1/4$ is normalization factor. In order to determine this state, Alice and Bob
perform joint measurement to obtain mean value of
observables $\sigma_\alpha \otimes \sigma_\beta$, which is equal to the
factor $c_{\alpha\beta}$ in principle
 $c_{\alpha\beta} = {\rm Tr} \left[ \rho^{AB} \sigma_\alpha \otimes \sigma_\beta
 \right]$.
Thus, through joint measurements Alice and Bob can determine such bipartite
quantum states.

Here, we give a general protocol based on the principle of maximal
entropy~\cite{PhysRev.106.620, PhysRev.108.171} for this LBQST task. The protocol here
is for anyonic SSR, and can be directly generalized to the relevant version for other
SSR. To distinguish between conventional quantum states $\rho$ and anyonic states $\tilde{\rho}$,
we use an additional tilde to denote anyonic state.
For bipartite anyonic states $\tilde{\rho}^{AB}$, the procedure for Alice and Bob
to determine $\tilde{\rho}^{AB}$ can be summarized as follows:

(1) Alice and Bob prepare a complete set of observables for their own subsystems
\begin{equation}
    \left\{ M_{c, i_c}^{A/B} ~|~ c \in {\rm ob}(\mathcal{C}), ~
        i_c =0, 1,\cdots, (N_c^{A/B})^2 - 1 \right\}, \nonumber
\end{equation}
where the superscript $A/B$ denotes subsystem $A$ or $B$, and $c$ is topological charge
in some anyon model $\mathcal{C}$.
For each sector labeled by $c$, there are $(N_c^{A/B})^2$ observables.
One $M_{c, 0_c}^{A/B}$ is the identity matrix in sector $c$, other $M_{c, i_c}^{A/B}$ (traceless)
form Lie algebra $\mathfrak{s}\mathfrak{u}(N_c^{A/B})$ of group SU($N_c^{A/B}$) satisfying
\begin{align}
    \label{eq:orthonormal}
    M_{a, i_a}^{A/B} M_{b, j_b}^{A/B} = & \frac{1}{N_a^{A/B}} \delta_{ab} \delta_{ij}
        I_{N_a^{A/B}} \nonumber \\
    & + \delta_{ab} \sum_k \left( i f_{i j k} + d_{i j k} \right) M_{a, k_a}^{A/B}.
\end{align}

(2) Alice and Bob perform joint measurements
\begin{equation}
    \tilde{\rm Tr} \left[ \tilde{\rho}^{AB} M^A_{a, i_a} \otimes M^B_{b, j_b} \right]
        = p_{i_a, j_b},~~~ \forall~ a, b, i_a, j_b, \nonumber
\end{equation}
where $\tilde{\rm Tr}$ is quantum trace~\cite{bonderson2017anyonic}.

(3) Alice and Bob build a set $\mathcal{Q}(\tilde{\rho}^{AB})$ which consists all
anyonic states $\tilde{\sigma}$ meet
\begin{equation}
    \mathcal{Q}(\tilde{\rho}^{AB}) = \left\{ \tilde{\sigma}  |  \tilde{\rm Tr}
        \left[ \tilde{\sigma} M^A_{a, i_a} \otimes M^B_{b, j_b} \right] = p_{i_a j_b},
        \forall ~ a, b, i_a, j_b \right\}. \nonumber
\end{equation}
The anyonic state inferred through LBQST task is the state with maximal entropy among
set $\mathcal{Q}(\tilde{\rho}^{AB})$,
i.e., $ \tilde{\sigma}_{m}(\tilde{\rho}^{AB}) = {\rm argmax}~ \tilde{S}(\tilde{\sigma})$,
where $\tilde{S}(\tilde{\sigma}) = - \tilde{\rm Tr} \left[ \tilde{\sigma}
 {\rm log}_2 \tilde{\sigma} \right]$
is the anyonic von Neumann entropy~\cite{bonderson2017anyonic}.

Using this protocol, we have following theorem.

\emph{Theorem} 1.--
For any bipartite quantum state $\rho^{AB}$ of quantum system with no SSR,
we can determine it through protocol given above,
i.e., $\sigma_m(\rho^{AB}) = \rho^{AB}$, and we have
 $\sigma_m (\rho^{AB}) = \sum_{i,j} p_{ij} M_i^A \otimes M_j^B$.

\emph{Proof}.--
The proof is straightforward. Since quantum state $\rho^{AB}$ with no superselection
rules can be decomposed into the form as:
 $\rho^{AB} = \sum_{i,j} c_{ij} M_i^A \otimes M_j^B$,
where $c_{ij}$ are unknown coefficients. Thus, taking advantage of
Eq.~(\ref{eq:orthonormal}), Alice and Bob can determine it through joint measurements.
\hfill$\blacksquare$

However, it is not true for Alice and Bob to be able to accurately determine any
anyonic bipartite states through LBQST task. It can be seen from the fact that
the Hilbert space of anyonic system $AB$ is not the tensor product of subsystems $A$
and $B$, and it is dominated by superselection rules.
Using this protocol, Alice and Bob can be able to determine the anyonic states of
their subsystems, however they cannot completely determined the whole bipartite
anyonic states in most cases. Since the state $\tilde{\sigma}$ in anyonic system $AB$
has general form
\begin{align}
    \label{eq:anyonicstate}
    \tilde{\sigma} = \sum_{c} \sum_{a, a', b, b', \vec{m}, \vec{m}', \vec{n}, \vec{n}'}
    \alpha_{c, a, a', b, b', \vec{m}, \vec{m}', \vec{n}, \vec{n}'}
    \begin{tikzpicture}[baseline, scale = 0.8]
        \draw (-0.5,0.25) -- (0.5,-0.25) node[pos = 0.5, above]{$c$};
        \draw (-0.5,0.75) -- (-0.5,0) node[pos = 0.2, left]{$a$};
        \draw (0.5,0.75) -- (0.5,0) node[pos = 0.2, right]{$b$};
        \draw (-0.5,-0.75) -- (-0.5,0) node[pos = 0.2, left]{$a'$};
        \draw (0.5,-0.75) -- (0.5,0) node[pos = 0.2, right]{$b'$};
        \draw (-0.9,1.15) -- (-0.5,0.75) -- (-0.1,1.15);
        \draw (-0.5,0.8) node[above]{$\vec{m}$};
        \draw (0.1,1.15) -- (0.5,0.75) -- (0.9,1.15);
        \draw (0.5,0.8) node[above]{$\vec{n}$};
        \draw (-0.9,-1.15) -- (-0.5,-0.75) -- (-0.1,-1.15);
        \draw (-0.5,-0.8) node[below]{$\vec{m}'$};
        \draw (0.9,-1.15) -- (0.5,-0.75) -- (0.1,-1.15);
        \draw (0.5,-0.8) node[below]{$\vec{n}'$};
    \end{tikzpicture}, \nonumber
\end{align}
where vectors $\vec{m}$ and $\vec{m}'$ denote the variables in subsystem $A$,
vectors $\vec{n}$ and $\vec{n}'$ denote the variables in subsystem $B$,
and $\alpha_{c, a, a', b, b', \vec{m}, \vec{m}', \vec{n}, \vec{n}'}$ are coefficients.
The state $\tilde{\sigma}$ has terms that cannot be represented by the tensor product
of local observables, only those terms where $c = 1$ can.
Thus, Alice and Bob were unable to learn about some information
in anyonic system. This inaccessible information used to determine the anyonic state
is hidden in the correlations, which is a non-local property depends on
superselection rules in anyonic system and fusion rules of anyon models. This kind
of correlations, which we call topological correlation, can be defined
as
\begin{equation}
    \label{eq:topocor}
    C_{topo}(\tilde{\rho}^{AB}) = \tilde{S}(\tilde{\sigma}_m(\tilde{\rho}^{AB}))
        - \tilde{S}(\tilde{\rho}^{AB}).
\end{equation}
The topological correlation is nonnegative by definition. It is obvious that
conventional bipartite quantum states without superselection rules and
the inferred anyonic states $\tilde{\sigma}_m(\tilde{\rho}^{AB})$ of any anyonic
state $\tilde{\rho}^{AB}$ have no topological correlation.

In order to obtain the topological correlation of anyonic bipartite
state $\tilde{\rho}^{AB}$, we should know the inferred
state $\tilde{\sigma}_m(\tilde{\rho}^{AB})$ through LBQST task, and it is an optimization
problem. However, we have the following important theorem which gives the analytic
form of the inferred state $\tilde{\sigma}_m(\tilde{\rho}^{AB})$.

\emph{Theorem} 2.--
For a given anyonic bipartite state $\tilde{\rho}^{AB}$ with maximal rank, its inferred
state $\tilde{\sigma}_m(\tilde{\rho}^{AB})$ through LBQST task can be expressed as
\begin{equation}
    \tilde{\sigma}_m(\tilde{\rho}^{AB}) = \sum_{a, b, i_a, j_b} p_{i_a, j_b}
        \frac{1}{d_a d_b} M^A_{a, i_a} \otimes M^B_{b, j_b},
\end{equation}
where $1/d_a d_b$ are normalization coefficients.

This theorem can be proved by the method of Lagrange multipliers (See
Appendix~\ref{section:app} for details). Using this theorem, we can calculate the
topological correlation of anyonic states with maximal rank analytically.
Let's consider the following anyonic state of four Fibonacci anyons with fusion
rule $\tau \times \tau = 1 + \tau$, where $\tau$ is Fibonacci anyon,
\begin{widetext}
    \begin{align}
        \tilde{\rho}^{AB}_{4\tau} = & p_1 \frac{1}{d_\tau^2}~
            \begin{tikzpicture}[baseline, scale = 0.8]
                \draw (-0.75,0.5) -- (-0.5,0.25) -- (-0.25,0.5);
                \draw (0.75,0.5) -- (0.5,0.25) -- (0.25,0.5);
                \draw (-0.75,-0.5) -- (-0.5,-0.25) -- (-0.25,-0.5);
                \draw (0.75,-0.5) -- (0.5,-0.25) -- (0.25,-0.5);
            \end{tikzpicture} + p_2 \frac{1}{d_\tau^{5/2}}~
            \begin{tikzpicture}[baseline, scale = 0.8]
                \draw (-0.75,0.5) -- (-0.5,0.25) -- (-0.25,0.5);
                \draw (0.75,0.5) -- (0.5,0.25) -- (0.25,0.5);
                \draw (-0.75,-0.5) -- (-0.5,-0.25) -- (-0.25,-0.5);
                \draw (0.75,-0.5) -- (0.5,-0.25) -- (0.25,-0.5);
                \draw (-0.5,0.25) -- (-0.5,-0.25);
            \end{tikzpicture} + p_3 \frac{1}{d_\tau^{5/2}}~
            \begin{tikzpicture}[baseline, scale = 0.8]
                \draw (-0.75,0.5) -- (-0.5,0.25) -- (-0.25,0.5);
                \draw (0.75,0.5) -- (0.5,0.25) -- (0.25,0.5);
                \draw (-0.75,-0.5) -- (-0.5,-0.25) -- (-0.25,-0.5);
                \draw (0.75,-0.5) -- (0.5,-0.25) -- (0.25,-0.5);
                \draw (0.5,0.25) -- (0.5,-0.25);
            \end{tikzpicture} + p_4 \frac{1}{d_\tau^2}~
            \begin{tikzpicture}[baseline, scale = 0.8]
                \draw (-0.75,1) -- (-0.5,0.75) -- (-0.25,1);
                \draw (0.75,1) -- (0.5,0.75) -- (0.25,1);
                \draw (-0.75,-1) -- (-0.5,-0.75) -- (-0.25,-1);
                \draw (0.75,-1) -- (0.5,-0.75) -- (0.25,-1);
                \draw (-0.5,0.75) -- (0,0.25) -- (0.5,0.75);
                \draw (-0.5,-0.75) -- (0,-0.25) -- (0.5,-0.75);              
            \end{tikzpicture} + p_5 \frac{1}{d_\tau^{5/2}}~
            \begin{tikzpicture}[baseline, scale = 0.8]
                \draw (-0.75,1) -- (-0.5,0.75) -- (-0.25,1);
                \draw (0.75,1) -- (0.5,0.75) -- (0.25,1);
                \draw (-0.75,-1) -- (-0.5,-0.75) -- (-0.25,-1);
                \draw (0.75,-1) -- (0.5,-0.75) -- (0.25,-1);
                \draw (-0.5,0.75) -- (0,0.25) -- (0.5,0.75);
                \draw (-0.5,-0.75) -- (0,-0.25) -- (0.5,-0.75);
                \draw (0,0.25) -- (0,-0.25);
            \end{tikzpicture}, \nonumber
    \end{align}
\end{widetext}
where black line denotes Fibonacci anyon, and $p_i$ for $i=1,\cdots,5$
are positive real number satisfying $\sum_{i=1}^5 p_i =1$. Using Theorem~2, we have
\begin{align}
    \tilde{\sigma}_m(\tilde{\rho}^{AB}_{4\tau})
    = & p_1 M^A_{1,0} \otimes M^B_{1,0} \nonumber \\
    & + p_2 \frac{1}{d_\tau} M^A_{\tau, 0} \otimes M^B_{1, 0}
    + p_3 \frac{1}{d_\tau} M^A_{1,0} \otimes M^B_{\tau, 0} \nonumber \\
    & + (p_4 + p_5) \frac{1}{d_\tau^2} M^A_{\tau, 0} \otimes M^B_{\tau, 0}, \nonumber
\end{align}
where operators $M_{1,0} = \ket{\tau, \tau; 1} \bra{\tau, \tau; 1}$
and $M_{\tau, 0} = \ket{\tau, \tau; \tau} \bra{\tau, \tau; \tau}$.
Thus, the topological correlation in anyonic state $\tilde{\rho}^{AB}_{4\tau}$
is $C_{topo}(\tilde{\rho}^{AB}_{4\tau}) = p_4 {\rm log}_2 \frac{p_4 d_\tau^2}{p_4 + p_5}
 + p_5 {\rm log}_2 \frac{p_5 d_\tau}{p_4 + p_5}$.
It can be prove that $C_{topo}(\tilde{\rho}^{AB}_{4\tau}) \ge 0$,
and when $p_4 = p_5/d_\tau$, the topological
correlations $C_{topo}(\tilde{\rho}^{AB}_{4\tau})$ reaches it minimum value.
Since that, when $p_4 = p_5/d_\tau$, we have
\begin{align}
    & p_4 \frac{1}{d_\tau^2}~
            \begin{tikzpicture}[baseline, scale = 0.8]
                \draw (-0.75,1) -- (-0.5,0.75) -- (-0.25,1);
                \draw (0.75,1) -- (0.5,0.75) -- (0.25,1);
                \draw (-0.75,-1) -- (-0.5,-0.75) -- (-0.25,-1);
                \draw (0.75,-1) -- (0.5,-0.75) -- (0.25,-1);
                \draw (-0.5,0.75) -- (0,0.25) -- (0.5,0.75);
                \draw (-0.5,-0.75) -- (0,-0.25) -- (0.5,-0.75);              
            \end{tikzpicture} + p_4 \frac{1}{d_\tau^{3/2}}~
            \begin{tikzpicture}[baseline, scale = 0.8]
                \draw (-0.75,1) -- (-0.5,0.75) -- (-0.25,1);
                \draw (0.75,1) -- (0.5,0.75) -- (0.25,1);
                \draw (-0.75,-1) -- (-0.5,-0.75) -- (-0.25,-1);
                \draw (0.75,-1) -- (0.5,-0.75) -- (0.25,-1);
                \draw (-0.5,0.75) -- (0,0.25) -- (0.5,0.75);
                \draw (-0.5,-0.75) -- (0,-0.25) -- (0.5,-0.75);
                \draw (0,0.25) -- (0,-0.25);
            \end{tikzpicture}
            = p_4 M^A_{\tau, 0} \otimes M^B_{\tau, 0}. \nonumber
\end{align}
This anyonic state $\tilde{\rho}^{AB}_{4\tau}$
can be determined by LOCC, i.e., $\tilde{\sigma}_m(\tilde{\rho}^{AB}_{4\tau})
 = \tilde{\rho}^{AB}_{4\tau}$.

Theorem~2 requires quantum states to be with maximal rank, while most states
have non-maximal rank. Luckily, according to Ref.~\cite{PhysRevLett.101.180505},
we can deal with states without maximal rank by treating them as the limit of
a series of states with maximal rank. For example, any anyonic state $\tilde{\rho}$
is the limit $p \rightarrow 0$ of anyonic states, $\tilde{\rho}(p)
 = (1-p)~ \tilde{\rho} + p~ \tilde{\rho}_m$,
where $\tilde{\rho}_m$ is the maximum mixed anyonic state in the same anyonic Hilbert
space as state $\tilde{\rho}$. It can be seen that $\tilde{\rho}(p)$ is anyonic state
with maximal rank as long as $p \neq 0$. Thus, the inferred state of $\tilde{\rho}$ can
be obtained by $\tilde{\sigma}(\tilde{\rho})
 = \lim_{p \rightarrow 0} \tilde{\sigma}(\tilde{\rho}(p))$.
Using the above method we can obtain the topological correlation of pure state
\begin{align}
    \ket{\psi} = \sqrt{q} \frac{1}{d_\tau}~
        \begin{tikzpicture}[baseline, scale = 0.8]
            \draw (-0.75,0.25) -- (-0.5,0) -- (-0.25,0.25);
            \draw (0.75,0.25) -- (0.5,0) -- (0.25,0.25);
        \end{tikzpicture} + \sqrt{1-q} \frac{1}{d_\tau}~
        \begin{tikzpicture}[baseline, scale = 0.8]
            \draw (-0.75,0.25) -- (-0.5,0) -- (-0.25,0.25);
            \draw (0.75,0.25) -- (0.5,0) -- (0.25,0.25);
            \draw (-0.5,0) -- (0,-0.5) -- (0.5,0);
        \end{tikzpicture}, \nonumber
\end{align}
where $q \in (0,1)$. The corresponding inferred state is $\tilde{\sigma}(\ket{\psi})
 = q M^A_{1, 0} \otimes M^B_{1, 0}
 + (1-q) \frac{1}{d_\tau^2} M^A_{\tau, 0} \otimes M^B_{\tau, 0}$.
Thus, the topological correlation is $C_{topo}(\ket{\psi})
 = H(q, 1-q) + (1-q) {\rm log}_2d_\tau^2$,
where $H(q, 1-q) = -q{\rm log}_2q + (1-q){\rm log}_2(1-q)$ is the binary entropy.
When $q=1/\mathcal{D}^2$, the topological correlation reaches its maximum
value $2{\rm log}_2\mathcal{D}$, which is double as TEE,
where $\mathcal{D} = \sqrt{1+d_\tau^2}$ is the total quantum dimension of
Fibonacci anyon model. This two-fold relationship can be easily seen from a
quantum information perspective. Since the correlations is twice as much as entanglement
entropy for pure states and the topological correlation defined here is
inaccessible information due to the anyonic SSR. Therefore, the entanglement
entropy of anyonic pure states will be reduced by half topological correlation,
numerically. This part is only a qualitative discussion. For more rigorous elaboration,
we will prove that topological correlation is equivalent to entropy of
anyonic charge entanglement in the following part. The latter can theoretically
derive TEE~\cite{bonderson2017anyonic}.

\emph{Topological correlation and entropy of anyonic charge entanglement}--
In this part we investigate the relation between the topological
correlation defined here and the entropy of anyonic charge entanglement(ACE)
defined in Ref.~\cite{bonderson2017anyonic}, and find that they describe
the same correlation.

First we review the entropy of anyonic charge entanglement.
For a given bipartite anyonic state $\tilde{\rho}$, the ACE can be defined as
\begin{equation}
    \label{eq:ace}
    C_{ACE}(\tilde{\rho}) = \tilde{S} \left( D_{A:B}[\tilde{\rho}] \right)
        - \tilde{S}\left( \tilde{\rho} \right),
\end{equation}
where $D_{A:B}$ is the superoperator severing charge lines that connect
two subsystem $A$ and $B$:
\begin{align}
    D_{A:B} \left[ \begin{tikzpicture}[baseline, scale = 0.7]
        \draw (-0.5,-0.75) node[below]{$a'$} -- (-0.5,0.75) node[above]{$a$};
        \draw (0.5,-0.75) node[below]{$b'$} -- (0.5,0.75) node[above]{$b$};
        \draw (0.5,-0.2) -- (-0.5,0.2) node[pos = 0.5, above]{$c$};
    \end{tikzpicture}
    \right] = \delta_{a,a'} \delta_{b,b'} \delta_{c,1}
    \begin{tikzpicture}[baseline, scale =0.7]
        \draw (-0.5,-0.75) -- (-0.5,0.75) node[above]{$a$};
        \draw (0.5,-0.75)  -- (0.5,0.75) node[above]{$b$};
        \draw (0,0) node[]{$\otimes$};
    \end{tikzpicture}, \nonumber
\end{align}
where we have suppressed variables within subsystems $A$ and $B$,
and showed the variables in fusion space connecting these two subsystems.
By acting such superoperator $D_{A:B}$, anyonic state $\tilde{\rho}$
will be mapped into space $D_{A:B}[\tilde{\rho}] \in \left( \bigoplus_a V^A_a \right)
 \otimes \left( \bigoplus_b V^B_b \right)$,
where $V^{A/B}_a$ is the linear operator space of subsystem $A$ or $B$
with total charge $a$. For such states with maximal rank, we can always
determine it through LBQST task using Theorem~2,
i.e., $\tilde{\sigma}_m(D_{A:B}[\tilde{\rho}]) = D_{A:B}[\tilde{\rho}]$.
Now, we have a theorem:

\emph{Theorem 3}.-- For a given bipartite anyonic state $\tilde{\rho}$
with maximal rank, its topological correlation is equal to its entropy of
anyonic charge
\begin{equation}
    C_{topo}(\tilde{\rho}) = C_{ACE}(\tilde{\rho}).
\end{equation}

\emph{Proof}-- To show that topological correlation is the same as
ACE, we need to show state $D_{A:B}[\tilde{\rho}]$ is the same
as $\tilde{\sigma}_m(\tilde{\rho})$ by noticing Eqs.~(\ref{eq:topocor})
and (\ref{eq:ace}). We are going to prove it in three steps. The first step is
to prove $D_{A:B}[\tilde{\rho}]$ is the state in set $\mathcal{Q}(\tilde{\rho})$.
The second step is to explain that $\tilde{\sigma}_m(D_{A:B}[\tilde{\rho}])
= \tilde{\sigma}_m(\tilde{\rho})$ by definition.
Finally, since $\tilde{\rho}$ is anyonic state with maximal rank,
we have $\tilde{\sigma}_m(D_{A:B}[\tilde{\rho}]) = D_{A:B}[\tilde{\rho}]$
($D_{A:B}[\tilde{\rho}]$ must be state with maximal rank),
then,  we obtain $D_{A:B}[\tilde{\rho}] = \tilde{\sigma}_m(\tilde{\rho})$.

The key to proof is the first step.
We will see that states $\tilde{\rho}$ and $D_{A:B}[\tilde{\rho}]$ give the same
result of joint measurement.
To this end, we are going to check that the general terms
\begin{align}
    K = & \ket{(\vec{m};a)_A, (\vec{n};b)_B; c} \bra{(\vec{m}';a)_A, (\vec{n}';b)_B; c} \nonumber\\
    = & \left( \frac{d_c^2}{d_{\vec{m}} d_{\vec{n}} d_{\vec{m}'} d_{\vec{n}}'} \right)^{\frac{1}{4}}
    \begin{tikzpicture}[baseline, scale = 0.8]
        \draw (0,0.25) -- (0,-0.25) node[pos = 0.5, right]{$c$};
        \draw (-0.5,0.75) -- (0,0.25) node[pos = 0.2, below]{$a$};
        \draw (0.5,0.75) -- (0,0.25) node[pos = 0.2, below]{$b$};
        \draw (-0.5,-0.75) -- (0,-0.25) node[pos = 0.2, above]{$a'$};
        \draw (0.5,-0.75) -- (0,-0.25) node[pos = 0.2, above]{$b'$};
        \draw (-0.9,1.15) -- (-0.5,0.75) -- (-0.1,1.15);
        \draw (-0.5,0.8) node[above]{$\vec{m}$};
        \draw (0.1,1.15) -- (0.5,0.75) -- (0.9,1.15);
        \draw (0.5,0.8) node[above]{$\vec{n}$};
        \draw (-0.9,-1.15) -- (-0.5,-0.75) -- (-0.1,-1.15);
        \draw (-0.5,-0.8) node[below]{$\vec{m}'$};
        \draw (0.9,-1.15) -- (0.5,-0.75) -- (0.1,-1.15);
        \draw (0.5,-0.8) node[below]{$\vec{n}'$};
    \end{tikzpicture},\nonumber
\end{align}
where the symbol $({d_a}/{d_{\vec{m}}})^{1/4}$ is used to briefly
represent the normalization factor of the quantum state $\ket{\vec{m};a}$,
in anyonic state $\tilde{\rho}$ satisfy this relation.
Suppose $\ket{\vec{p}; e}_A \bra{\vec{p}'; e} \otimes \ket{\vec{q}; d}_B \bra{\vec{q}'; d}$
is the general term in measured
space $\left( \bigoplus_a V^A_a \right) \otimes \left( \bigoplus_b V^B_b \right)$,
then, we are going to check the equation:
\begin{align}
    \label{eq:target}
    & \tilde{\rm Tr} \left[ \ket{\vec{p}; e}_A \bra{\vec{p}'; e} \otimes
        \ket{\vec{q}; f}_B \bra{\vec{q}'; f} K \right] \nonumber \\
    = & \tilde{\rm Tr} \left[ \ket{\vec{p}; e}_A \bra{\vec{p}'; e} \otimes
        \ket{\vec{q}; f}_B \bra{\vec{q}'; f} D_{A:B}[K] \right].
\end{align}
By using $F$ move,
\begin{equation}
    \begin{tikzpicture}[baseline, scale = 0.7]
        \draw (0,0.25) -- (0,-0.25) node[pos = 0.5, right]{$c$};
        \draw (-0.5,0.75) -- (0,0.25) node[pos = 0, above]{$a$};
        \draw (0.5,0.75) -- (0,0.25) node[pos = 0, above]{$b$};
        \draw (-0.5,-0.75) -- (0,-0.25) node[pos = 0, below]{$a'$};
        \draw (0.5,-0.75) -- (0,-0.25) node[pos = 0, below]{$b'$};
    \end{tikzpicture} = \sum_g \left[ F^{ab}_{a'b'} \right]_{cg}
    \begin{tikzpicture}[baseline, scale = 0.7]
        \draw (-0.5,-0.75) node[below]{$a'$} -- (-0.5,0.75) node[above]{$a$};
        \draw (0.5,-0.75) node[below]{$b'$} -- (0.5,0.75) node[above]{$b$};
        \draw (0.5,-0.2) -- (-0.5,0.2) node[pos = 0.5, above]{$g$};
    \end{tikzpicture}, \nonumber
\end{equation}
where $\left[ F^{ab}_{a'b'} \right]_{cg}$ is $F$ matrix,
and $\left[ F^{ab}_{ab} \right]_{c1} = \sqrt{d_c/d_a d_b}$, we get the same result
for the left and right sides of the Eq.~(\ref{eq:target}),
 $\delta_{a,a'} \delta_{b,b'} \delta_{e,a} \delta_{f,b}
 \delta_{\vec{m},\vec{p}'} \delta_{\vec{n},\vec{q}'} \delta_{\vec{m}',\vec{p}}
 \delta_{\vec{n}',\vec{q}} d_c$.
Thus, states $\tilde{\rho}$ and $D_{A:B}[\tilde{\rho}]$ give the same
result of joint measurement. Thus, Eq.~(\ref{eq:target}) holds,
and topological correlation is same as ACE. \hfill$\blacksquare$

ACE measures the correlations that come from the charge lines connecting
the total charges of two subsystems, which breaks the direct product structure of Hilbert
space, while topological correlation measures the information that cannot be obtained
from the space of the direct product of subsystems. Indeed, both actually describe
the same thing.

\emph{Summary}--
In this letter, we use information-theoretic approach based on the principle of
maximal entropy to investigate the topological correlation in anyonic bipartite
states. Our work extends the method
in Ref.~\cite{PhysRevLett.101.180505, PhysRevLett.89.207901} by adding classical
communications between local parties. This leads to the identification of bipartite
quantum states without SSR but not of anyonic bipartite states.
This inaccessible information in anyonic system is defined as topological
correlation, which we prove to be equivalent to the entropy of anyonic charge
entanglement defined in the previous literature. This letter not only gives the
analytic results of topological correlation in anyonic bipartite states with
maximal rank, but also gives an operational meaning to this entropy of anyonic
charge entanglement which is formulated to explain topological entanglement entropy
in many-body system.
Although we have only discussed bipartite systems here, our work is illuminating
and can be easily generalized to multipartite systems.
We hope our work might shed light on anyonic quantum
information and Long-range entangled topological phases.

\begin{acknowledgments}
    This work is supported by National Key Research and Development Program of
    China (Grant No. 2021YFA0718302 and No. 2021YFA1402104), National Natural Science
    Foundation of China (Grants No. 12075310), and the Strategic Priority Research
    Program of Chinese Academy of Sciences (Grant No. XDB28000000).
\end{acknowledgments}

\appendix

\onecolumngrid

\section{The proof of Theorem~2}\label{section:app}

In this section, we aim to prove the Theorem~2 in main text.
For a given anyonic state $\tilde{\rho}^{AB}$,
we are going to find the anyonic state $\tilde{\sigma}_m(\tilde{\rho}^{AB})$ with the
maximal anyonic von Neumann entropy among the set $\mathcal{Q}(\tilde{\rho}^{AB})$
which consists all anyonic states $\tilde{\sigma}$ satisfying
\begin{equation}
    \label{eq:conditions}
    \tilde{\rm Tr} \left[ \tilde{\sigma} M^A_{a, i_a} \otimes M^B_{b, j_b} \right]
        = \tilde{\rm Tr} \left[ \tilde{\rho}^{AB} M^A_{a, i_a} \otimes M^B_{b, j_b}
        \right],~~~~~ \forall~ a, b, i_a, j_b,
\end{equation}
where $M^A_{a, i_a}$ are $i_a$-th observables with trace zero in $N^A_a$-dimensional
subsystem $A$ with total charge $a$. One $M^{A/B}_{a, 0_a}$ is identity matrix
in sector $a$, others $M^{A/B}_{a, i_a}$ for $i \neq 0$ form
Lie algebra $\mathfrak{s}\mathfrak{u}(N^A_a)$ of group SU($N^A_a$).
And $\tilde{\sigma}$ is the anyonic  state of system $AB$, which has the form
\begin{align}
    \tilde{\sigma} = \sum_{c} \sum_{a, a', b, b', \vec{m}, \vec{m}', \vec{n}, \vec{n}'}
    \alpha_{c, a, a', b, b', \vec{m}, \vec{m}', \vec{n}, \vec{n}'}
    \begin{tikzpicture}[baseline, scale = 0.8]
        \draw (-0.5,0.25) -- (0.5,-0.25) node[pos = 0.5, above]{$c$};
        \draw (-0.5,0.75) -- (-0.5,0) node[pos = 0.2, left]{$a$};
        \draw (0.5,0.75) -- (0.5,0) node[pos = 0.2, right]{$b$};
        \draw (-0.5,-0.75) -- (-0.5,0) node[pos = 0.2, left]{$a'$};
        \draw (0.5,-0.75) -- (0.5,0) node[pos = 0.2, right]{$b'$};
        \draw (-0.9,1.15) -- (-0.5,0.75) -- (-0.1,1.15);
        \draw (-0.5,0.8) node[above]{$\vec{m}$};
        \draw (0.1,1.15) -- (0.5,0.75) -- (0.9,1.15);
        \draw (0.5,0.8) node[above]{$\vec{n}$};
        \draw (-0.9,-1.15) -- (-0.5,-0.75) -- (-0.1,-1.15);
        \draw (-0.5,-0.8) node[below]{$\vec{m}'$};
        \draw (0.9,-1.15) -- (0.5,-0.75) -- (0.1,-1.15);
        \draw (0.5,-0.8) node[below]{$\vec{n}'$};
    \end{tikzpicture}, \nonumber
\end{align}
where vectors $\vec{m}$ and $\vec{m}'$ denote the variables in subsystem $A$,
vectors $\vec{n}$ and $\vec{n}'$ denote the variables in subsystem $B$,
and $\alpha_{c, a, a', b, b', \vec{m}, \vec{m}', \vec{n}, \vec{n}'}$ are coefficients.

We prove Theorem~2 by the method of Lagrange multipliers. Specifically, we try to find
the minimum value of the following formula
\begin{equation}
    - \tilde{S}(\tilde{\sigma}) - \sum_{a, b, i_a, j_b} \Lambda_{i_a, j_b}
        \left( \tilde{\rm Tr} \left[ \tilde{\sigma} M^A_{a, i_a} \otimes M^B_{b, j_b}
        \right] - p_{i_a, j_b} \right),
\end{equation}
where $\tilde{S}(\tilde{\sigma}) = - \tilde{\rm Tr} \left[ \tilde{\sigma}
{\rm log}_2 \tilde{\sigma} \right]$ is anyonic von Neumann entropy,
$\Lambda_{i_a, j_b}$ are the Lagrange multipliers, and $p_{i_a, j_b}
 = \tilde{\rm Tr} \left[ \tilde{\rho}^{AB} M^A_{a, i_a} \otimes M^B_{b, j_b} \right]$.
Then, we have
\begin{align}
        & - \tilde{S}(\tilde{\sigma}) - \sum_{a, b, i_a, j_b} \Lambda_{i_a, j_b} \left( \tilde{\rm Tr} \left[ \tilde{\sigma}
            M^A_{a, i_a} \otimes M^B_{b, j_b} \right] - p_{i_a, j_b} \right) \nonumber \\
        = & \tilde{\rm Tr} \left[ \tilde{\sigma} ~ {\rm ln}~ \tilde{\sigma} \right]
            - \sum_{a, b, i_a, j_b} \tilde{\rm Tr} \left[ \tilde{\sigma} ~ {\rm ln}~
            e^{\Lambda_{i_a, j_b} M^A_{a, i_a} \otimes M^B_{b, j_b}}  \right]
            + \sum_{a, b, i_a, j_b} \Lambda_{i_a, j_b} p_{i_a, j_b} \nonumber \\
        = & \tilde{\rm Tr} \left[ \tilde{\sigma} ~ \left( {\rm ln}~ \tilde{\sigma}
            -  \sum_{a, b, i_a, j_b} {\rm ln}~ e^{\Lambda_{i_a, j_b} M^A_{a, i_a}
            \otimes M^B_{b, j_b}} \right) \right]
            + \sum_{a, b, i_a, j_b} \Lambda_{i_a, j_b} p_{i_a, j_b} \nonumber \\
        = & \tilde{\rm Tr} \left[ \tilde{\sigma} ~ \left( {\rm ln}~ \tilde{\sigma}
            -  {\rm ln}~ \prod_{a, b, i_a, j_b} e^{\Lambda_{i_a, j_b} M^A_{a, i_a}
            \otimes M^B_{b, j_b}} \right) \right]
            + \sum_{a, b, i_a, j_b} \Lambda_{i_a, j_b} p_{i_a, j_b} \nonumber \\
        \ge & \tilde{\rm Tr} \left[  \tilde{\sigma}
            -  \prod_{a, b, i_a, j_b} e^{\Lambda_{i_a, j_b} M^A_{a, i_a} \otimes M^B_{b, j_b}} \right]
        + \sum_{a, b, i_a, j_b} \Lambda_{i_a, j_b} p_{i_a, j_b}.
\end{align}
In the last step, we have used the anyonic Klein inequality~\cite{bonderson2017anyonic},
 $\tilde{\rm Tr} \left[ A \left( {\rm ln} ~A - {\rm ln}~B \right) \right]
 \ge \tilde{\rm Tr} \left[ A - B \right]$, for positive definite anyonic operators $A$
and $B$, where the equality is satisfied
if and only if $A = B$. In the other word,
when $\tilde{\sigma} = \prod_{a, b, i_a, j_b} e^{\Lambda_{i_a, j_b} M^A_{a, i_a}
 \otimes M^B_{b, j_b}}$, anyonic von Neumann entropy $\tilde{S}(\tilde{\sigma})$
reaches its maximum. Then, we have
\begin{equation}
    \label{eq:a4}
    \tilde{\sigma}_m(\tilde{\rho}^{AB}) = \prod_{a, b, i_a, j_b} e^{\Lambda_{i_a, j_b} M^A_{a, i_a}
    \otimes M^B_{b, j_b}},
\end{equation}
where Lagrange multipliers $\Lambda_{i_a, j_b}$ can be determined by
conditions~(\ref{eq:conditions}). 

By using formula
\begin{align}
    M_{a, i_a}^{A/B} M_{b, j_b}^{A/B} = & \frac{1}{N_a^{A/B}} \delta_{ab} \delta_{ij}
    I_{N_a^{A/B}} \nonumber \\
& + \delta_{ab} \sum_k \left( i f_{i j k} + d_{i j k} \right) M_{a, k_a}^{A/B}.
\end{align}
where $f_{ijk}$ and $d_{ijk}$ are the structure constants, $N^{A/B}_a$ is the dimension
of subsystem $A$ or $B$ with total charge $a$, we can reformulate
Eq.~(\ref{eq:a4}) as
\begin{equation}
    \tilde{\sigma}_m(\tilde{\rho}^{AB}) = \sum_{a, b, i_a, j_b} c_{i_a, j_b}
        M^A_{a, i_a} \otimes M^B_{b, j_b},
\end{equation}
where coefficients $c_{i_a, j_b}$ are functions of Lagrange
multipliers $\Lambda_{i_a, j_b}$, which can also be determined by
conditions~(\ref{eq:conditions}).
Thus, we have
\begin{equation}
    \tilde{\sigma}_m(\tilde{\rho}^{AB}) = \sum_{a, b, i_a, j_b} p_{i_a, j_b}
    \frac{1}{d_a d_b} M^A_{a, i_a} \otimes M^B_{b, j_b},
\end{equation}
where $1/d_a d_b$ is normalization coefficient due to the fact
that $\tilde{\rm Tr}[I_{N^A_a}] = N^A_a d_a$. That's the result of the Theorem~2
in main text. It should be noted that we have used the anyonic Klein inequality
where positive definite anyonic operators are required. Thus, we should
limit ourselves to the anyonic states $\tilde{\rho}^{AB}$ with maximal rank.
\hfill$\blacksquare$

\twocolumngrid

\bibliographystyle{apsrev4-2} \bibliography{AQDbib.bib}

\end{document}